\documentclass[journal]{IEEEtran}
\usepackage{graphicx}
\graphicspath{{suman/}}
\usepackage{hyperref,graphicx}
\usepackage{amsmath}
\usepackage{color}
\usepackage{nopageno}
\usepackage{multicol}

\begin{document}
\title{Electron-Acoustic Solitary Waves in Fermi Plasma With Two-Temperature Electrons}

\author{S. Pramanick, A. Dey and S. Chandra
\thanks{S. Pramanick is with the Department of Physics, Indian Institute of Technology Kharagpur, Kharagpur, West Bengal 721302, India (e-mail: sumanhorse@gmail.com; sumanhorse@iitkgp.ac.in)}
\thanks{A. Dey is with Department of Physics, Lady Brabourne College, Kolkata 700017, West Bengal, India (e-mail: ankitaankidey@hotmail.com; ankitadey179@gmail.com) }

\thanks{S. Chandra is with Govt. General Degree College at Kushmandi, Dakshin Dinajpur, 733121, India. (e-mail: swarniv147@gmail.com)}}

\maketitle
\begin{abstract}
Electron Acoustic waves in Fermi Plasma with two temperature electrons have various applications in space and laboratory-made plasma. In some dense plasma systems like the inside of compact stars, Fermi plasma is important. We have studied Fermi plasma system with three components, two temperature electrons, and ions. The hot electrons are mobile and produce restoring force to the system while cold electrons are immobile and produce inertia to the system. We have studied the dispersion behavior of electron acoustic waves in Fermi plasma with two temperature electrons and investigated its dependence with various plasma parameters. We have investigated KdV-B equation for the solitary profile of Fermi plasma with two temperature electrons and investigated its dependence with various plasma parameters.
\end{abstract}

\section{Introduction}
\IEEEPARstart{P}{}lasma systems containing two distinct groups of electrons shows Electron acoustic waves (EAWs). The distinction between two electron groups comes from their energy. Two types of electrons are 1) Hot electrons and 2) Cold electrons. The frequency of electron acoustic mode is higher than ion acoustic frequencies of plasmas. Hot electrons can freely move with less viscous drag and supply restoring force, whereas cold electron feels viscous drag and produce inertia to the system. The thermal speed of hot electron is very large in comparison to cold electron. The phase speed of electron acoustic wave is smaller than the thermal speed of hot electrons but much larger than thermal speed of cold electrons. In this system ions may be considered as uniform neutralizing background. EAWs with two groups of electrons plays an important role in space plasma (Ang and Zhang (2007)\cite{ref17}; Barnes et al. (2003)\cite{ref20}; Feldman et al. (1975)\cite{ref30}, (1983a)\cite{ref31} as well as laboratory made plasmas (Ditmire et al. (1998)\cite{ref27}; Sheridan et al. (1991)\cite{ref50}; Armstrong et al. (1979)\cite{ref18}; Kadomtsev and Pogutse (1971)\cite{ref37}; Henry and Trguier (1972)\cite{ref36}; Defler and Simonen (1969)\cite{ref25}). The source of broadband electrostatic noises can be addressed with EAWs and it has been used to explain those. The wave emission in different regions of earth’s atmosphere can be explained with EAWs. For these studies EAWs has become one of the important research areas in plasma physics. In recent years there is a boost in the study of the nonlinear evolution of EAWs (Bains et al. (2011)\cite{ref19}; Soultana and Kourakis (2011)\cite{ref55}; Kourakis and Shukla (2004)\cite{ref39}; Singh and Lakhina (2001)\cite{ref54}). Various space-craft missions, e,g, the FAST at the auroral region (Ergun et al. (1998a\cite{ref28}, 1998b\cite{ref29}); Delory et al. (1998)\cite{ref26}; Pottelette et al. (1999)\cite{ref46}) and the POLAR and GEOTAIL missions in the magnetosphere (Matsumoto et al. (1994)\cite{ref44}; Franz et al. (1998)\cite{ref33}; Cattell et al. (2003)\cite{ref23}) explained by EAW related structures. Most of these application sites need theory and better understanding of non-relativistic classical plasmas. However, there are numbers of works on the theory of nonlinear propagation of electrostatic modes in quantum plasmas with consideration of quantum hydrodynamic model of plasma (Shukla and Eliasson (2006)\cite{ref52}; Sahu and Roychoudhury (2006)\cite{ref48}; Ali and Shukla (2006)\cite{ref51}; Manfredi (2005)\cite{ref41}; Haas et al. (2003)\cite{ref35}; Gardner and Ringhofer (1996)\cite{ref40}). The non-linear wave structure of cold and hot electrons are investigated \cite{ref1}-\cite{ref7} \\
In this paper we studied linear and nonlinear properties of EAWs in Fermi plasma with two temperature electrons. We first assumed basic hydrodynamic equations for the system then we normalized them using suitable scaling for the system. With the normalized equations on hand we linearized them to get linear dispersion relation and linear dispersion characteristics are investigated. The Korteweg-de Vries (KdV) equation is derived using standard perturbation technique with taking only odd powers of perturbation fraction. We investigated dependence of soliton properties on different parameters of plasma. Then we conclude the paper with some remarks and future work plans.


\section{Basic Formulation}\label{BE}
The plasma system that we considered is unmagnetized consisting two groups of electrons. Two groups are 1) Hot electrons and 2) Cold electrons. The thermal energy of hot electron is higher than cold electrons so the mobility of hot electron is large compared to that of cold electrons. So, in the momentum equation for hot electron we consider inertia term as zero, whereas in the momentum equation of cold electrons we consider a viscous term as its mobility is less and more responsive to the viscous forces.
\subsubsection{Continuity equation}
\begin{equation}\label{eq1}
\frac{\partial n_{h}}{\partial t}+\frac{\partial\left(n_{h} u_{h}\right)}{\partial x}=0
\end{equation}
\begin{equation}\label{eq2}
\frac{\partial n_{c}}{\partial t}+\frac{\partial\left(n_{c} u_{c}\right)}{\partial x}=0
\end{equation}

\subsubsection{Momentum equation}
\begin{equation}\label{eq3}
0=\frac{1}{m_{e}}\left[e \frac{\partial \phi}{\partial x}-\frac{1}{n_{h}} \frac{\partial P_{h}}{\partial x}+\frac{\hbar^{2}}{2 m_{e}} \frac{\partial}{\partial x}\left[\frac{1}{\sqrt{n_{h}}} \frac{\partial^{2} \sqrt{n_{h}}}{\partial x^{2}}\right]\right]
\end{equation}

This is momentum equation for hot electrons. The inertia term (left hand side of the equation) is zero, as hot electrons are very mobile so its inertia can be assumed to be zero.

\begin{multline}\label{eq4}
m_e \left(\frac{\partial}{\partial t}+u_{c} \frac{\partial}{\partial x}\right) u_{c} =e \frac{\partial \phi}{\partial x}
+\frac{\hbar^{2}}{2 m_{e}} \frac{\partial}{\partial x}\left[\frac{1}{\sqrt{n_{c}}} \frac{\partial^{2} \sqrt{n_{c}}}{\partial x^{2}}\right]\\
+\eta_{c} \frac{\partial^{2} u_{c}}{\partial x^{2}}
\end{multline}

This is momentum equation for cold electrons. The inertia term is non zero as the mobility of cold electrons are very less. For our two-electron plasma system cold electron produce the restoring force for electron acoustic oscillations. Cold electrons also got a viscous term.

\subsubsection{Poisson's equation}
\begin{equation}\label{eq5}
\frac{\partial^{2} \phi}{\partial x^{2}}=4 \pi e \left(n_{c}+n_{h}-Z_{i} n_{i}\right)
\end{equation}

The subscript $i$ in the Poisson’s equation is for representing ions. The above-mentioned equations are the governing equations for the plasma system, where $n_h$ and $n_c$ are the densities of hot and cold electrons respectively. $\phi$ is the electrostatic potential, $P_h$ is the pressure law for hot electrons. $\hbar$ is the plank constant divided by $2\pi$, $m_e$ is the mass of electron and $e$ is the charge of electron, $Z_{i}e$ is the charge of ion and $\eta_c$ is the viscous constant for cold electrons. 

\subsubsection{The pressure law}
We considered Fermi pressure as the main pressure component for hot electrons. Fermi pressure is
\begin{equation}\label{eq6}
P_{j}=\frac{m_{j} V_{F j}^{2}}{3 n_{j 0}^{2}} n_{j}^{3} 
\end{equation}

where $j=h$ for hot electron, $j=c$ for cold electron and $j=i$ for ions, $m_{j}$ is the mass, $n_{j}$ is the number density of $j$ particles with equilibrium concentration $n_{j 0}, V_{F j}$ is the Fermi thermal speed which is given as $V_{F j}=\sqrt{2 k_{B} T_{F j} / m_{j}}, T_{F j}$ is the Fermi temperature and $k_{B}$ is the Boltzmann's constant. The parameters need to be normalized to get a good control to the equations and normalized equations are easy to work with. Normalization means we need to define some suitable scaling associated to our problem. With these scaling constants we can make our parameters dimensionless.
For our problem normalization has done with following way\\
$x \rightarrow x \omega_{j} / V_{F j}$; $t \rightarrow t \omega_{j}$; $\phi \rightarrow e \phi / 2 k_{B} T_{F j}$; $u_{j} \rightarrow u_{j} / V_{F j}$; $n_{j} \rightarrow n_{j} / n_{j 0}$ and $\eta_{c} \rightarrow \eta_{c} \omega_{j} / m_{e} V_{F j}^{2}$ where $\omega_{j}=\sqrt{4 \pi e n_{c 0} e^{2}/m_{e}}$ is the plasma oscillation frequency.\\
Changing the variables and parameters in Eqs. (\ref{eq1}-\ref{eq5})  we have Normalized Governing Equations:

\begin{equation}\label{eq7}
\frac{\partial n_{h}}{\partial t}+\frac{\partial\left(n_{h} u_{h}\right)}{\partial x}=0
\end{equation}
\begin{equation}\label{eq8}
\frac{\partial n_{c}}{\partial t}+\frac{\partial\left(n_{c} u_{c}\right)}{\partial x}=0
\end{equation}
\begin{equation}\label{eq9}
0=\frac{\partial \phi}{\partial x}-n_{h} \frac{\partial n_{h}}{\partial t}+\frac{H^{2}}{2} \frac{\partial}{\partial x}\left[\frac{1}{\sqrt{n_{h}}} \frac{\partial^{2} \sqrt{n_{h}}}{\partial x^{2}}\right]
\end{equation}

\begin{multline}\label{eq10}
\left(\frac{\partial}{\partial t}+u_{c} \frac{\partial}{\partial x}\right) u_{c}=\frac{\partial \phi}{\partial x}+\frac{H^{2}}{2} \frac{\partial}{\partial x}\left[\frac{1}{\sqrt{n_{c}}} \frac{\partial^{2} \sqrt{n_{c}}}{\partial x^{2}}\right]\\
+\eta_{c} \frac{\partial^{2} u_{c}}{\partial x^{2}}
\end{multline}
\begin{equation}\label{eq11}
\frac{\partial^{2}\phi}{\partial x^{2}}=n_{c}+\frac{n_{h}}{\delta}-\frac{\delta_{i}}{\delta} n_{i}
\end{equation}
where $H=\hbar \omega_{j} / 2 k_{B} T_{F j}$ is a non-dimensional quantum diffraction parameter, $\delta=n_{c 0} / n_{h 0}$ and $\delta_{i}=Z_{i}n_{i0}/n_{h0}$

\section{Dispersion Characteristics }\label{DR}

For investigating the nonlinearity in the behavior of electron acoustic wave, we assume following perturbation expansion for $n_h$, $n_c$, $u_h$, $u_c$ and $\phi$.

\begin{equation}\label{eq12}
\left[\begin{array}{c}
n_{h} \\
n_{c} \\
u_{h} \\
u_{c} \\
\phi
\end{array}\right]=\left[\begin{array}{c}
1 \\
1 \\
u_{0} \\
u_{0} \\
\phi_{0}
\end{array}\right]+\varepsilon\left[\begin{array}{c}
n_{h}^{(1)} \\
n_{c}^{(1)} \\
u_{h}^{(1)} \\
u_{c}^{(1)} \\
\phi^{(1)}
\end{array}\right]+\varepsilon^{3}\left[\begin{array}{c}
n_{h}^{(2)} \\
n_{c}^{(2)} \\
u_{h}^{(2)} \\
u_{c}^{(2)} \\
\phi^{(2)}
\end{array}\right]+\cdots
\end{equation}

Here we assume existence of streaming equilibrium velocity for hot and cold electrons and equilibrium field in constant field $\phi_0$. Substituting these expansions (Eq. \ref{eq12}) in the governing relations (Eqs. \ref{eq7}-\ref{eq11}) and then taking only linear terms (linearizing) with the assumption that all field variable varies periodically as $e^{i(k x-\omega t)}$, where $k$ is the wavenumber and $\omega$ is wave frequency, we have following complex dispersion relation:

\begin{equation}\label{eq13}
-k^2=\frac{1}{\left[\frac{{H}^{2}{k}^{2}}{4}-\left(\frac{\omega-u_0 k}{k}\right)^2\right]+i \eta_c\left(\omega-u_0 k\right)}+\frac{1/\delta}{1+\frac{{H}^2{k}^2}{4}} 
\end{equation}
 
 Now we have to keep in mind that $k$ itself is a complex number, $k=k_1+i k_2$. Putting this into the expression of Complex dispersion relation we have two equations, one for real part of the equation and other for imaginary part of the equation. The real dispersion equation is the solution for EAW of our three component plasma, is:
 
\begin{equation}\label{eq14}
    -1=\frac{A+\frac{B}{\delta}}{AB-C\eta_c}
\end{equation}
Where,
\begin{equation}
    A=\left[k_{1}^{2}+\frac{H^{2}k_{1}^{2}}{4}\right]
\end{equation}
\begin{equation}
    B=\left[H^{2}k_{1}^{2}-\omega^{2}-u_{0}^{2}k_{1}^{2}+2{\omega}u_{0}k_{1}\right]
\end{equation}
\begin{equation}
    C=k_{1}^{2}\left(\omega-k_{1}u_{0}\right)
\end{equation}

 With $k_2=0$, this is the linear dispersion relation of the EAW. If we plot $k_1$ vs $\omega$ then the plot will give us dispersion plot. $H$, $u_0$, $\eta_c$, $\delta$ are the parameters of the equation.
 
 \begin{figure}[ht]
{	\centering
	\includegraphics[width=3in,angle=0]{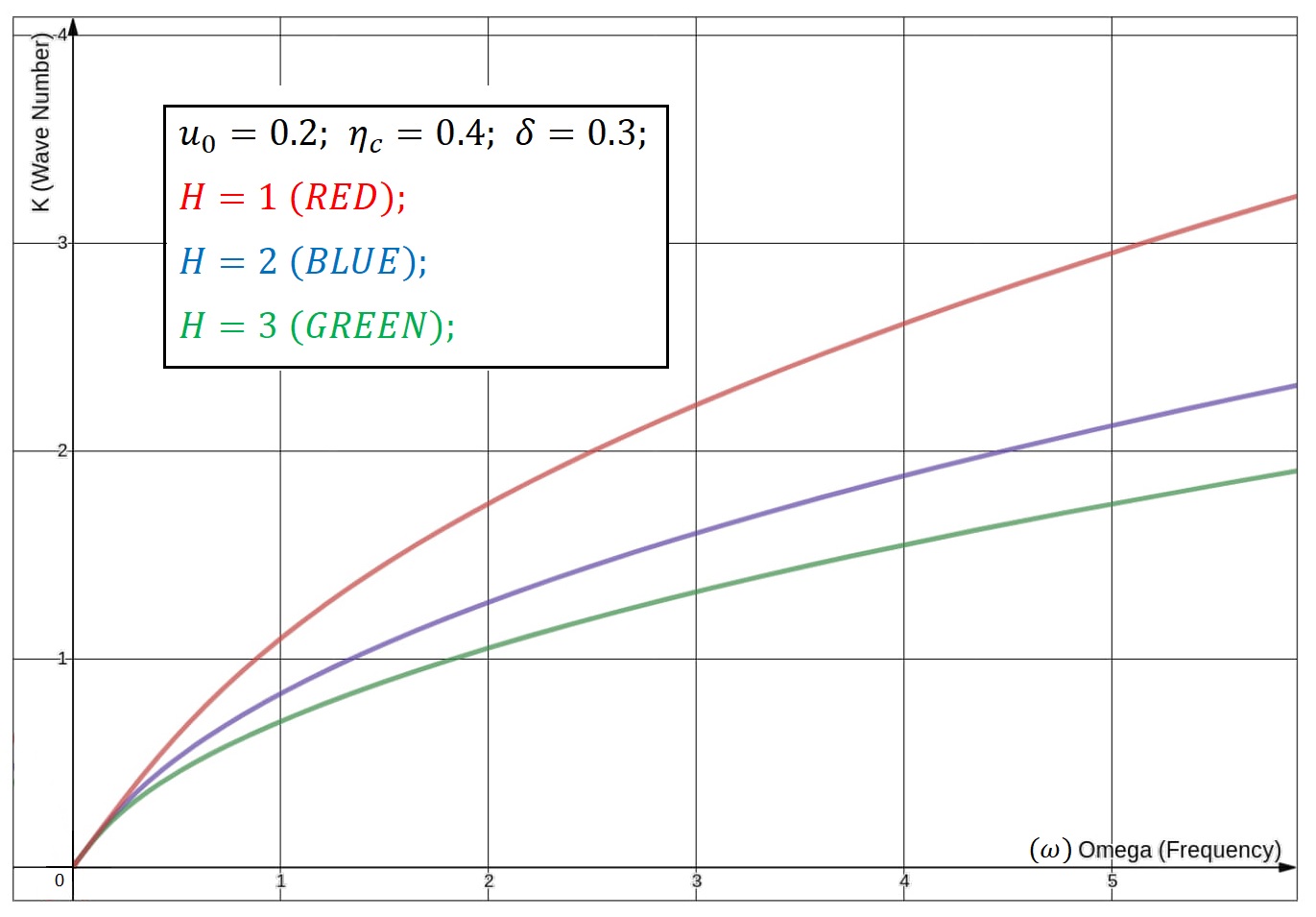}
	\caption{Real dispersion relation for different quantum diffraction parameter (H) keeping $u_{0}$, $\delta$, $\eta_c$ constant}
	\label{fig1}
}
\end{figure}

\begin{figure}[ht]
{	\centering
	\includegraphics[width=3in,angle=0]{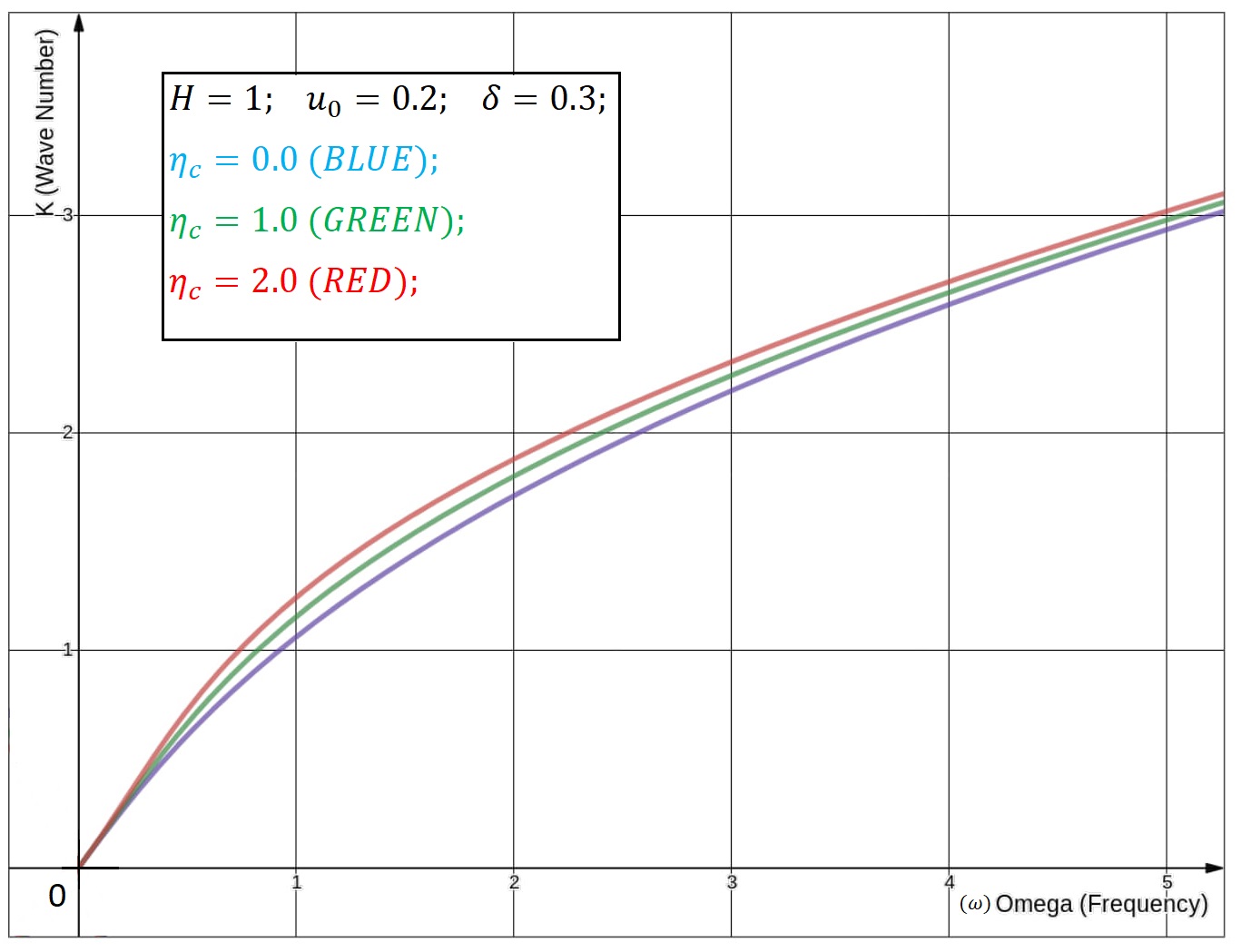}
	\caption{Real dispersion relation for different values of coefficient of viscosity $\eta_c$ keeping $u_{0}$, $H$, $\delta$ constant}
	\label{fig2}
}
\end{figure}

\begin{figure}[ht]
{	\centering
	\includegraphics[width=3in,angle=0]{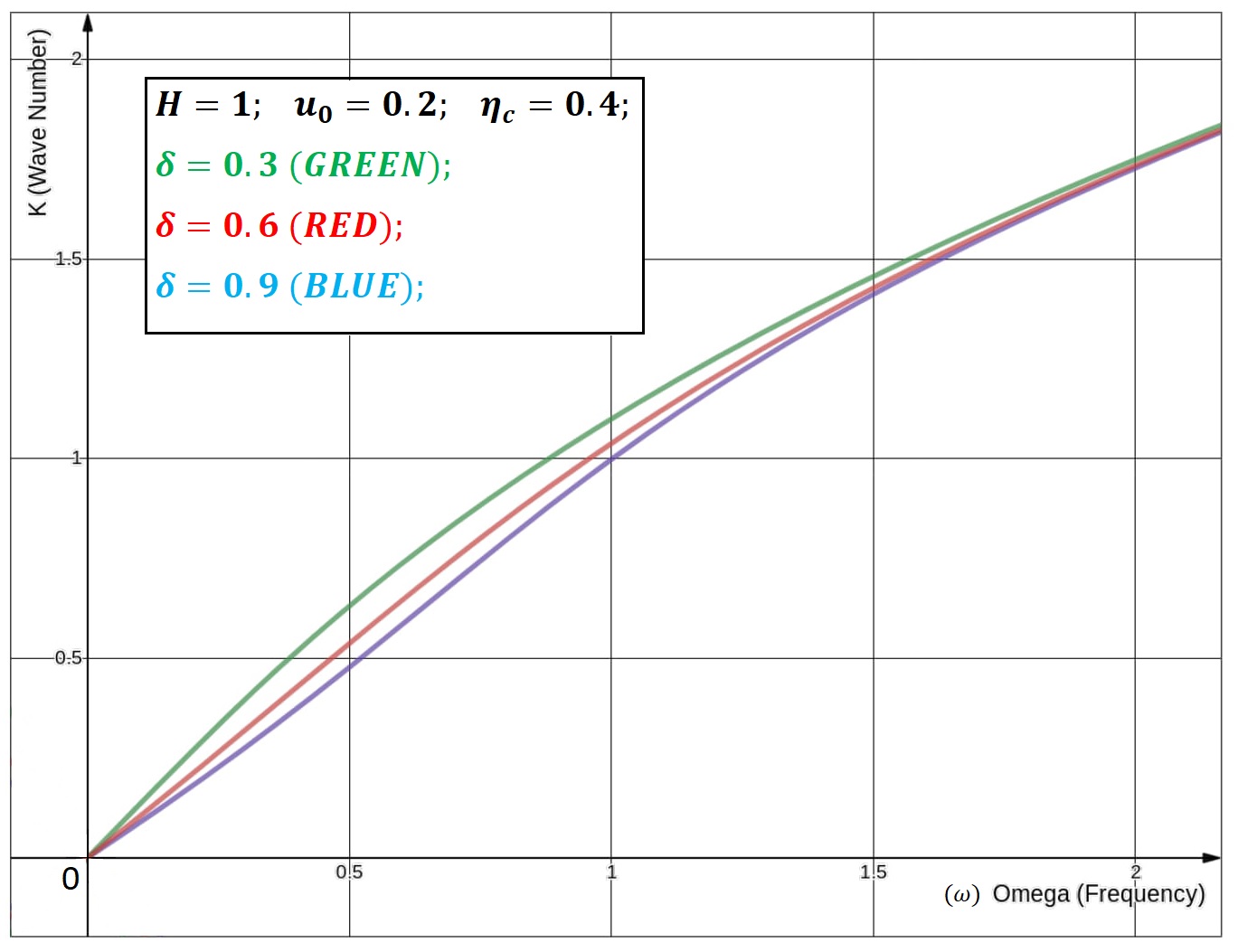}
	\caption{Real dispersion relation for different values of $\delta$ keeping  $\eta_c$, $u_{0}$ and $H$ constant}
	\label{fig3}
}
\end{figure}

\begin{figure}[ht]
{	\centering
	\includegraphics[width=3in,angle=0]{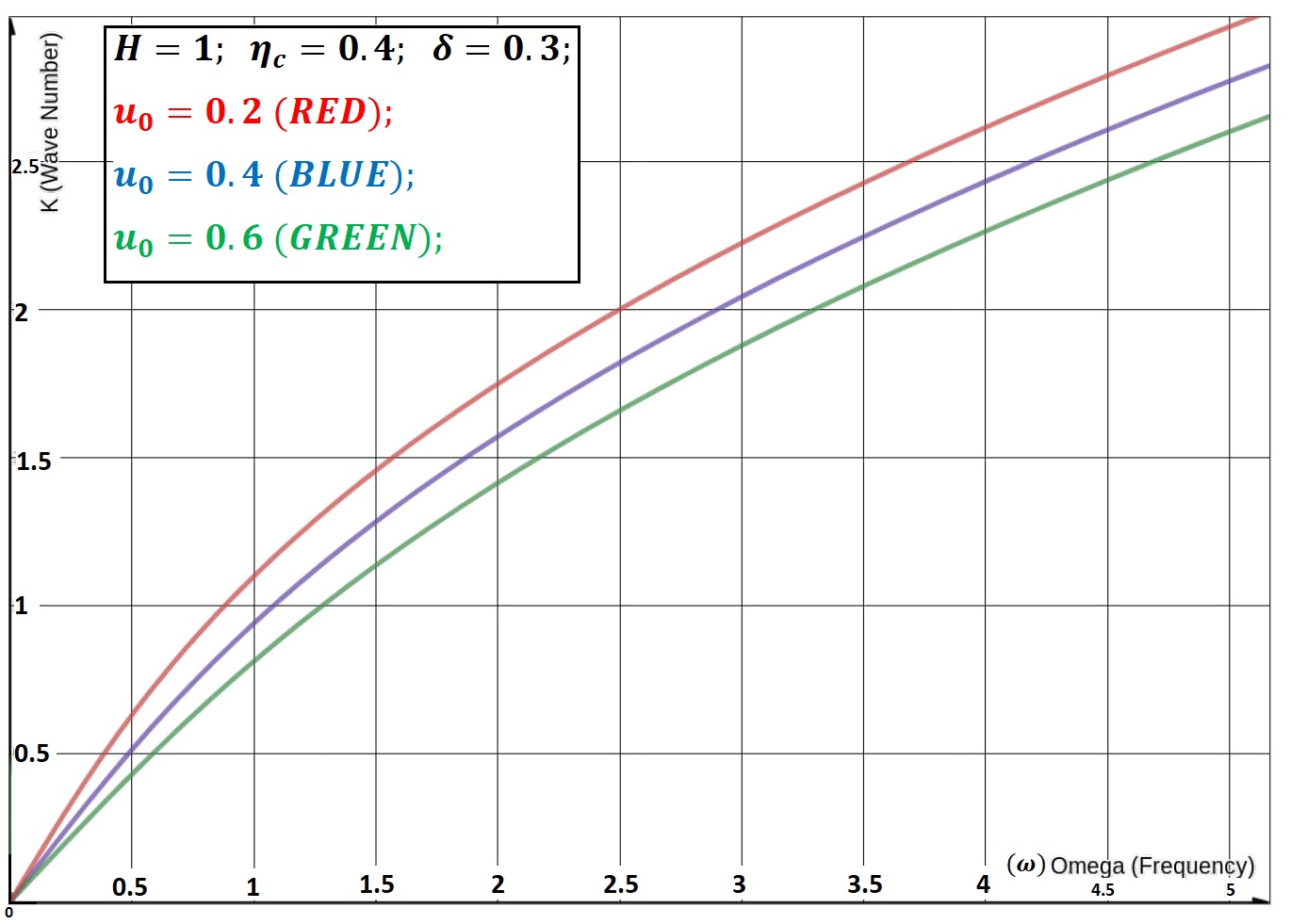}
	\caption{Real dispersion relation for different values of $u_0$ keeping  $\eta_c$, $\delta$ and $H$ constant}
	\label{fig4}
}
\end{figure}
 
 Fig, \ref{fig1}. shows $\omega$ vs $k$ plot for different $H$ values. We change $H$ from 1 to 3 keeping $u_0$, $\delta$ and $\eta_c$ constant. As $H$ increases the slope of $k$ vs $\omega$ plot decreases, that is also visible from the dispersion relation. However, $H$ depends on plasma frequency $\omega_c$ linearly. So, increasing $H$ means higher plasma frequency. Finally, we can conclude that for high plasma frequency the EAW has less wavenumber with same frequency.
 
 Fig. \ref{fig2}. shows dispersion plot for different values of $\eta_c$ keeping $H$, $u_0$ and $\delta$ constant. Plot shows that increase of $\eta_c$ results a increase in slope of $k$ vs $\omega$ plot. The slope of $k$ vs $\omega$ plot is inversely proportional to the EAW velocity in the plasma medium. So, an increase in viscous drag implies a decrease in the velocity of electron acoustic waves (EAW), which physically should be the case.

Fig. \ref{fig3}. shows dispersion plot for different values of $\delta$. Increase in $\delta$ shows decrease in slope (for small $\omega$ this behavior is clear). Increasing delta means more initial cold electrons compare to hot electrons. So, according to our plots a greater number of initial cold electrons implies a decrease in $k$ vs $\omega$ slop, that is increase in EAW velocity. Now, velocity of sound waves, rather acoustic waves increases if rigidity of the medium increases. A greater number of cold electrons means more viscous drag and eventually more rigid medium, which can be the cause for increase in EAW velocity.

Fig. \ref{fig4}. shows dispersion plot for different values of equilibrium streaming velocity $u_0$. Plot shows with increasing $u_0$ the slope of $k$ vs $\omega$ plot decreases, i.e. an increase in EAW velocity.

 \section{KdV-B Equation and It's Solution}\label{kdv-b}
\begin{figure}[ht]
{	\centering
	\includegraphics[width=\linewidth]{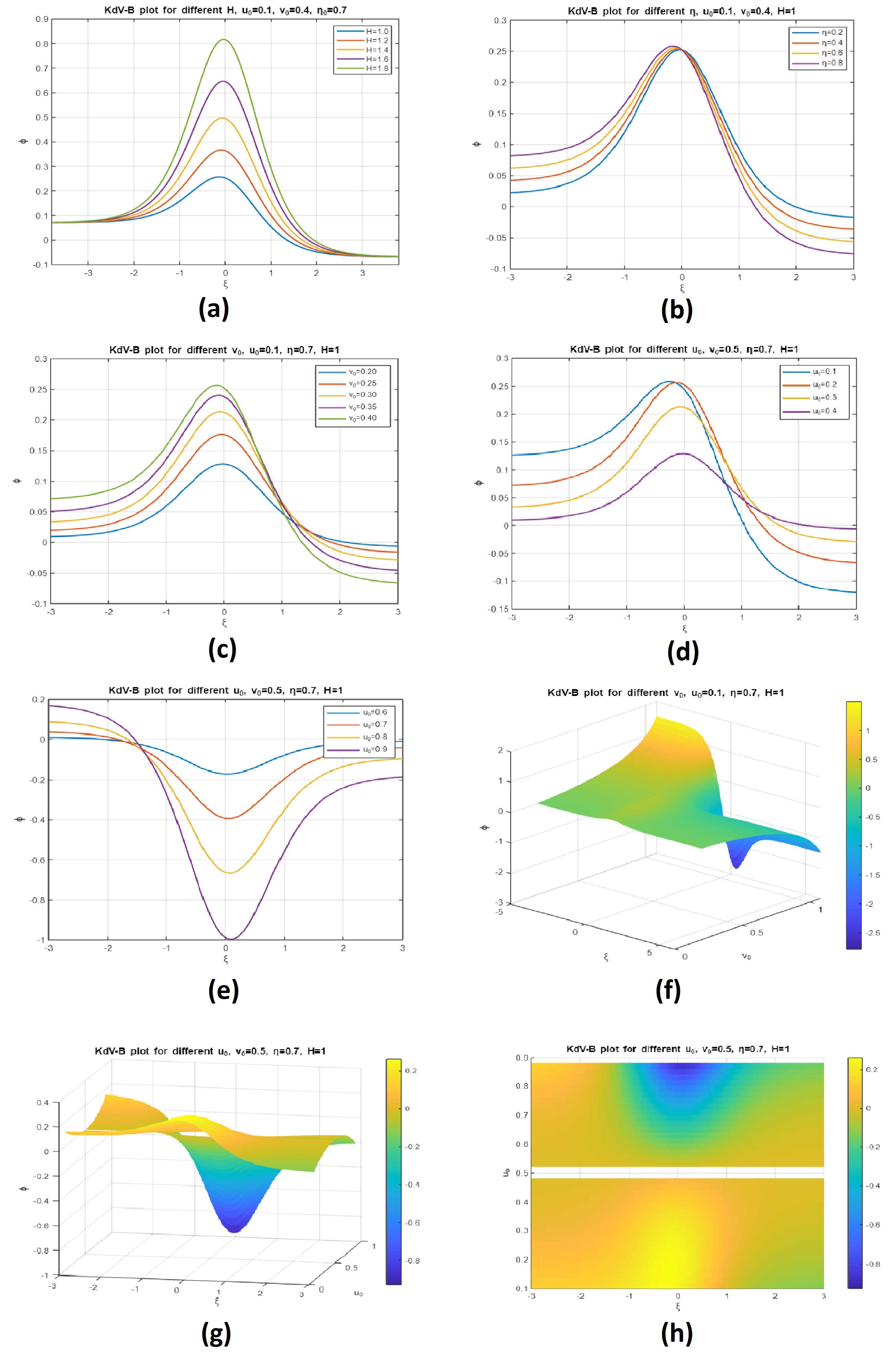}
	\caption{(a) KdV-B plot for different $H$ for fixed $u_0$, $v_0$ and $\eta_0$, (b) KdV-B plot for different $\eta_0$ for fixed $u_0$, $v_0$ and $H$, (c) KdV-B plot for different $v_0$ for fixed $u_0$, $H$ and $\eta_0$, (d) KdV-B plot for different $u_0$ keeping $H$, $v_0$ and $\eta_0$ constant, (e) KdV-B plot for different $u_0$ keeping $H$, $v_0$ and $\eta_0$ constant, (f) KdV-B three-dimensional plot for different $v_0$ for fixed $u_0$, $H$ and $\eta_0$, (g) KdV-B 3-D plot for different $u_0$ keeping $H$, $v_0$ and $\eta_0$ constant, (h) KdV-B color-plot for different $u_0$ keeping $H$, $v_0$ and $\eta_0$ constant}
	\label{fig5}
}
\end{figure}

\begin{figure}[ht]
{	\centering
	\includegraphics[width=\linewidth]{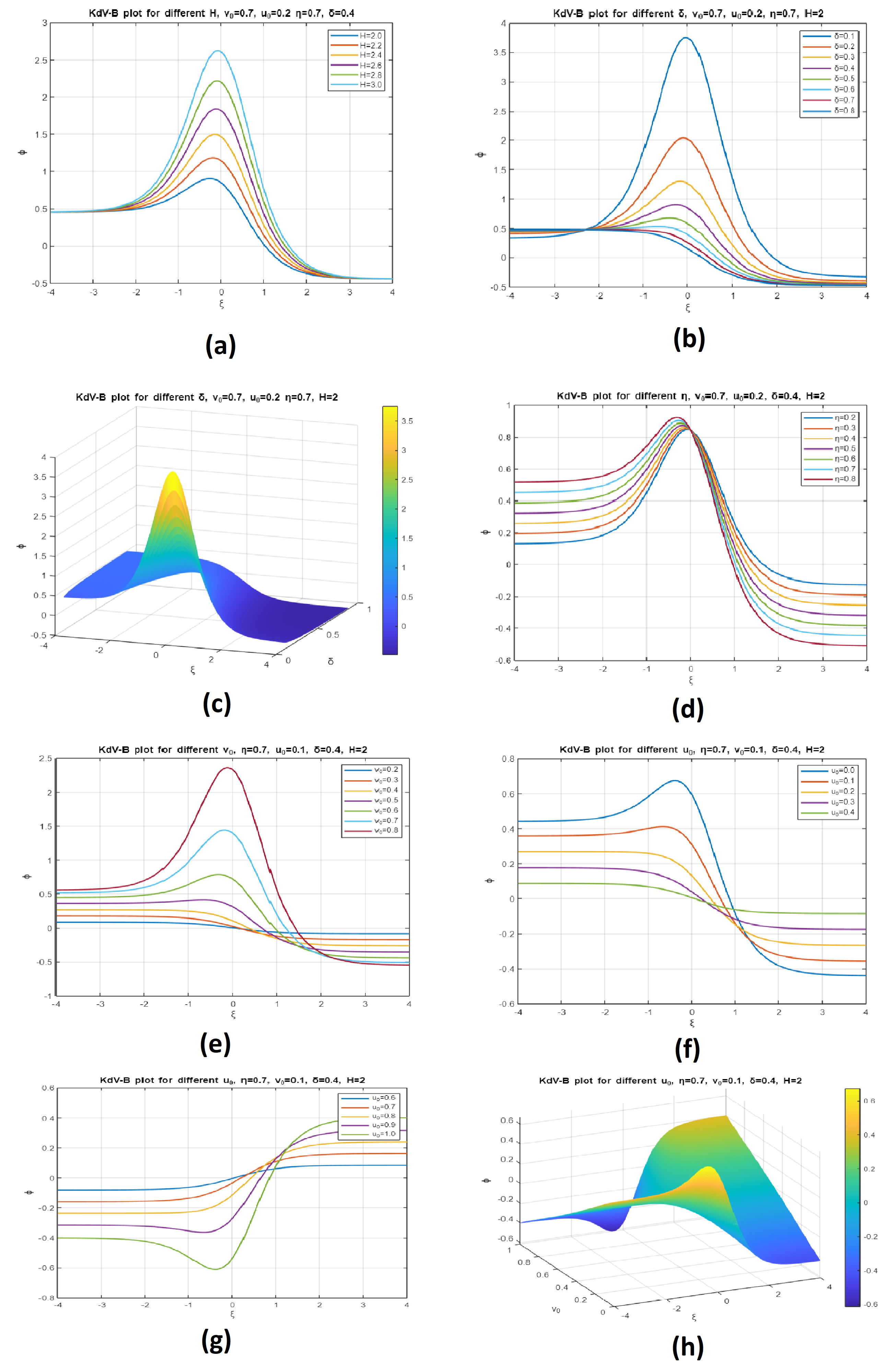}
	\caption{(a) KdV-B plot for different $H$ for fixed $u_0$, $v_0$, $\delta$ and $\eta_0$, (b) KdV-B plot for different $\delta$ for fixed $u_0$, $v_0$, $H$ and $\eta_0$, (c) KdV-B Surface plot for different $\delta$ for fixed $u_0$, $v_0$, $H$ and $\eta_0$, (d) KdV-B plot for different $\eta_0$ for fixed $u_0$, $v_0$, $H$ and $\delta$, (e) KdV-B plot for different $v_0$ for fixed $u_0$, $\eta_0$, $H$ and $\delta$, (f) KdV-B plot for different $u_0$ for fixed $v_0$, $\eta_0$, $H$ and $\delta$, (g) KdV-B plot for different $u_0$ for fixed $v_0$, $\eta_0$, $H$ and $\delta$, (h) KdV-B surface plot for different $u_0$ for fixed $v_0$, $\eta_0$, $H$ and $\delta$}
	\label{fig6}
}
\end{figure}

 We used standard reductive perturbation technique with usual stretching of space and time variables. We used two perturbation expansions to see the behavior of the solution of KdV-B equation. using the perturbation expansion Eq. \ref{eq12} and stretching of variables $\xi=\varepsilon^\frac{1}{2}\left(x-v_0 t\right)$; $\eta_c=\eta_0 \varepsilon^\frac{1}{2}$; $\tau=\varepsilon^\frac{3}{2} t$\\
 Putting them into governing equations and taking the lower order terms with power of epsilon we have the following KdV-Burger equation
 \begin{equation}\label{eq15}
 \frac{\partial \phi}{\partial \tau}+N\phi\frac{\partial \phi}{\partial \xi}+D\frac{\partial^3 \phi}{\partial \xi^3}-R\frac{\partial^2 \phi}{\partial \xi^2}=0
 \end{equation}
 with, $N=\frac{p_1^2\left(p_1+2p_2\right)-\left(p_1+p_2\right)}{2p_1\left(p_1+p_2\right)}$; $D=-\frac{H^2}{2}\frac{p_1+p_2+p_1 p_2}{p_1\left(p_1+p_2\right)}$; $R=\frac{p_1\eta_0}{2\left(p_1+p_2\right)}$
 where, $p_1=-\frac{1}{\left(v_0-u_0\right)}$ and $p_2=-\frac{1}{\left(v_0-u_0\right)^2}$
 for a perturbation of variables with all powers of $\varepsilon$:
\begin{equation}\label{eq16}
\left[\begin{array}{c}
n_{h} \\
n_{c} \\
u_{h} \\
u_{c} \\
\phi
\end{array}\right]=\left[\begin{array}{c}
1 \\
1 \\
u_{0} \\
u_{0} \\
\phi_{0}
\end{array}\right]+\varepsilon\left[\begin{array}{c}
n_{h}^{(1)} \\
n_{c}^{(1)} \\
u_{h}^{(1)} \\
u_{c}^{(1)} \\
\phi^{(1)}
\end{array}\right]+\varepsilon^{2}\left[\begin{array}{c}
n_{h}^{(2)} \\
n_{c}^{(2)} \\
u_{h}^{(2)} \\
u_{c}^{(2)} \\
\phi^{(2)}
\end{array}\right]+\cdots
\end{equation}
and stretching of variables
$\xi=\varepsilon^\frac{1}{2}\left(x-v_0 t\right)$; $\eta_c=\eta_0 \varepsilon^\frac{1}{2}$; $\tau=\varepsilon^\frac{3}{2}t$ 
we get another solution for KdV-B equation (\ref{eq15}) with
$N=p_1+\frac{v_0}{2}+\frac{v_0}{2p_2\delta}$; $D=\frac{V_0\left(H^2-4\delta\right)}{8p_2\delta}$; $R=-\frac{\eta_0}{2}$ with $p_1=-\frac{1}{\left(v_0-u_0\right)}$ and $p_2=-\frac{1}{\left(v_0-u_0\right)^2}$
The solution of KdV-B equation in asymptotic limit is
\begin{equation}\label{eq20}
    \phi=\frac{12D}{N}\left(1-\tanh^2{\xi}\right)-\frac{36R}{15N}\tanh{\xi}
\end{equation}
Fig. \ref{fig5} and Fig. \ref{fig6} shows $\phi$ vs $\xi$ plots for change in different parameters of the system. These plots are the solitary profiles of EAWs showing nonlinear behavior of the plasma system.

\section{Results and Discussions}\label{RD}

Fig.\ref{fig5}. (a) and Fig.\ref{fig6}. (a) shows that increase in height of solitary profile with increase in $H$ keeping $u_0$, $v_0$, $\delta$ and $\eta_0$ constant. Fig. \ref{fig5}. (b) and Fig. \ref{fig6}. (d) shows that with increase in the value of $\eta$ keeping other parameters constant, the difference of the field $(\phi)$ before and after soliton structure increases. Fig. \ref{fig5}. (c) and Fig. \ref{fig6}. (e) shows that with increase in $v_0$ keeping other parameters constant, the difference of the field $(\phi)$ before and after soliton structure increases. Fig. \ref{fig5}. (f) shows the behavior of soliton structure with full range change of $v_0$. Fig. \ref{fig5}. (d) and Fig. \ref{fig6}. (f) shows that with increase in $u_0$ keeping other parameters constant, the difference of the field $(\phi)$ before and after soliton structure decreases. In Fig. \ref{fig5}. (e) and Fig. \ref{fig6}. (g) the behavior just reversed, as for these plots the sign of $(v_0-u_0)$ changes. Fig. \ref{fig5}. (g), Fig. \ref{fig5}. (h) and Fig. \ref{fig6}. (h) shows change of solitary profile with full range change of $u_0$. Fig. \ref{fig5}. (g) and Fig. \ref{fig5}. (h) shows a discontinuity at $u_0=v_0$, which is at $u_0=0.5$. This is because at $u_0=v_0$ the solution is invalid as $p_1$ and $p_2$ both tend to infinity at this value. Fig. \ref{fig6}. (b) and Fig. \ref{fig6}. (c) shows that with increase in the value of $\delta$ keeping other plasma parameters constant the peak of solitary profile decreases and the difference of the field $(\phi)$ before and after soliton structure increases.

\section{Conclusions}\label{Conc}
Both the linear and nonlinear properties of electron-acoustic wave (EAWs) have been investigated in three component Fermi plasma consisting of two distinct groups of electrons and solitary ions. Dispersion relation that we obtained, is a general one including inertia effect for cold electrons and effect of Fermi pressure for hot electrons. The stable models for EAWs are shown. The dependence of electron-acoustic wave velocity on different plasma parameters has been shown and explained. The slope of the dispersion relation plots depends on the velocity of EAW inversely. The dependence of this EAW velocity and EAW frequency on different plasma parameters has been shown. Electron-acoustic wave velocity increases with increasing $H$, $u_0$ and $\delta$ and decreases with increasing viscous coefficient $\eta_c$.

To study nonlinear behavior of electron acoustic waves KdV-Burger’s equation has been derived. While deriving KdV-Burger’s equation in general perturbative reduction technique we assumed two kinds of perturbations, With usual perturbation containing terms with all integer powers of $\varepsilon$. and with a special kind of perturbation having only terms with odd powers of $\varepsilon$.The present investigation that we have done on this paper may be helpful in understanding the basic features of electron-acoustic waves in dense Fermi plasma systems like astrophysical objects like neutron stars, white dwarfs as well as laboratory-made Fermi plasma systems and laser-solid plasma experiments.

\bibliographystyle{IEEEtran}
\bibliography{suman}

\vfill

\end{document}